\documentstyle[12pt,preprint]{aastex}
\doublespace
\input epsf

\def\bib{\par\noindent\hangindent=3mm\hangafter=1}

\def\mv{M_V}

\def\msol{\mbox{M}_\odot}

\def\mmvol{(\mbox{M}_\odot)^{-1}\, \mbox{pc}^{-3}}

\def\beq{\begin{equation}}
\def\eeq{\end{equation}}

\def\simgr{\,\hbox{\hbox{$ > $}\kern -0.8em \lower 1.0ex\hbox{$\sim$}}\,}
\def\simle{\,\hbox{\hbox{$ < $}\kern -0.8em \lower 1.0ex\hbox{$\sim$}}\,}

\begin{document}

\title{The Galactic disk mass function: reconciliation of the HST and nearby determinations}

\author{Gilles Chabrier}
\affil{Ecole Normale Sup\'erieure de Lyon,\\
Centre de Recherche Astrophysique de Lyon (UMR CNRS 5574),
 69364 Lyon Cedex 07, France\\
(chabrier@ens-lyon.fr)}

\vskip 1.cm

\newpage

\begin{abstract}
We derive and parametrize the Galactic mass function (MF) below 1 $\msol$ characteristic of both single objects and
binary systems.
We resolve the long standing discrepancy between the MFs derived from the HST and from the
nearby luminosity functions, respectively. We show that this discrepancy stemmed from {\it two} cumulative effects, namely
(i) incorrect color-magnitude determined distances, due a substantial fraction of M dwarfs in the HST
sample belonging to the metal-depleted, thick-disk population, as corrected recently by Zheng et al. and (ii) unresolved binaries.
We show that both the nearby and HST MF for unresolved systems are consistent with a fraction $\sim $50\% of M-dwarf binaries, with the mass of both the primaries and
the companions originating from the same underlying single MF.
This implies that $\sim$30\%
of M dwarfs should have an M dwarf companion and  $\sim$20\% should have a brown dwarf companion,
in agreement with recent determinations. 
The present calculations show that the so-called "brown-dwarf desert" should be reinterpreted as a lack of high mass-ratio $(m_2/m_1\la 0.1$) systems, and does not preclude a substantial fraction of brown dwarfs as companions
of M dwarfs or for other brown dwarfs.
\end{abstract}

\section{Introduction}
\label{Intro}

The determination of the disk stellar luminosity function (LF) and mass function (MF) in the low-mass
star ($m\la 1\,\msol$) domain is still subject to debate and remains up to date an unsettled
issue.
The disagreement between the MF inferred from the photometric HST LF and from the nearby
5.2-pc LF has been a controversial issue since the Gould, Bahcall \& Flynn (1997; GBF97) paper.
The MF derived from the local sample keeps rising, although
moderately, down to the hydrogen burning limit, whereas the MF derived
from the HST LF is steadily decreasing from $0.6\, \msol$ down to $0.1\,\msol$
(see Figure 1 of M\'era, Chabrier and Schaeffer 1998).
The question is of prime importance for various reasons. First, the determination of the very shape of the MF
bears profound consequences for our understanding of star formation. Second,
whereas the luminosity of galaxies
arises mostly from stars from about 1 to a few solar masses,
most of their mass is contained in objects with
$m\le 1\,\msol$. The determination of the MF in the M-dwarf regime is thus crucial for a proper evaluation
of their mass budget and mass-to-light ratio. Third, the normalization of the MF near the hydrogen
burning limit is the cornerstone for an accurate evaluation of the brown dwarf (BD) content of the disk.
Last but not least, the M-dwarf present-day MF (PDMF)
represents the {\it initial } MF (IMF) of the Galaxy,
i.e. is representative of the mass distribution of the stars {\it ever} formed in the Galaxy (Scalo 1986), a central input in galactic evolution and cosmic star formation history.
The unresolved discrepancy between the HST and nearby MF determinations
thus prevents robust determinations of the aforementioned quantities.
In this Letter, we reconsider this problem in the light of the recently reanalyzed HST LF (Zheng et al. 2001).

\section{Initial mass function from the nearby sample}

The LF $\Phi(M)$
 requires the determination of the distance of the objects.
Samples with trigonometric parallax determination
require near distances from the Sun and define the so-called {\it nearby} LF $\Phi_{near}$.
A major advantage
of $\Phi_{near}$ is the identification of binary systems.
A V-band nearby LF can be derived by combining Hipparcos parallax data (ESA 1997)
for $\mv < 12$ and the sample of nearby stars 
with ground-based parallaxes (Dahn, Liebert \& Harington 1986) for $\mv>12$ to a completeness distance r=5.2 pc.
On the other hand,
Henry \& MCarthy (1990) used speckle interferometry to resolve companions of every known M dwarf within 5 pc
and obtained the complete M dwarf LF $\Phi_{near}$ in the H and K band. Their sample recovers the Dahn et al. (1986)
one, plus one previously unresolved companion (GL 866B).
Up to now, samples extended to a larger volume remain incomplete (see Henry et al. 1997)
and are hampered by ill-determined distances
(see Chabrier 2001, \S 3).

Kroupa (2001) and Chabrier (2001) have determined the Galactic disk M-dwarf MF from
the V-band 5-pc $\Phi_{near}$, although using different functional forms.
We have redone this analysis, by calculating the MF from both the aforementioned V-band 
and K-band $\Phi_{near}$.
Recently, Delfosse et al. (2000) and S\'egransan et al. (2002a) combined adaptative optics and accurate radial velocities to
determine the mass-magnitude relation of about 20 objects between $\sim 0.6$ and $\sim 0.09$ $\msol$ in the V, J, H and K bands with mass accuracies of 0.2 to 5\%.
The mass-magnitude relationships (MMR) derived from the Baraffe et al. (1998; BCAH98) models reproduce 
these data within less than 1-$\sigma$ in the J, H and K bands (Figure 3 of Delfosse et al. 2000).
The agreement is less good in the V band, with a systematic
offset of a few tenths of a magnitude below $\sim 0.3\,\msol$ ($\mv \ga 12$),
as discussed at length in BCAH98 and Chabrier et al. (2000, Figure 1).
The implications for the MF have been examined in detail by Chabrier (2001, Figures 1 and 2) and have been found to remain modest ($\la 15\%$ in the mass determination for $m\sim$0.2-0.3 $\msol$).
The theoretical M-dwarf radii of BACH98 also agree within 1\% or less for $m\le 0.5\,\msol$
with the radius measurements obtained recently with the VLTI by S\'egransan et al. (2002b).
This establishes the validity of deriving the MFs from the observed LFs using the theoretical BCAH98 MMRs.
However, in order to avoid any possible source of error, the conversion of the V-band LF into a MF was done using the Delfosse et al. (2000) $m$-$\mv$ relation, fitted to the data.
These results are displayed in Figure \ref{fig1_col.ps}. 
We note the
very good agreement between the two determinations, which establishes the consistency of the two
observed samples,
part of the $\sim$1.5-$\sigma$ difference in the mass range
$\log m \sim$ -0.5 to -0.6 reflecting most likely the remaining uncertainties in the MMR\footnote{The last bin is
very likely contaminated by young/massive BDs or still contracting very-low-mass stars with $m\la 0.12\,\msol$.
As shown in Chabrier (2002), an IMF including this bin extrapolated into the BD regime would
overestimate significantly the number of such objects.}.
The solid line displays a lognormal form which gives a fairly good representation of the results:

\begin{eqnarray}
\xi(\log \,m)={dn \over d\log \,m}=0.158\times \,\exp\{-{(\log \, m\,\,-\,\,\log \, 0.08)^2\over 2\times (0.69)^2}\}\,\,\,\,{\rm pc}^{-3}\,(\log \,\msol)^{-1}
\label{IMFdisk}
\end{eqnarray}

\noindent with the same normalization as Scalo (1986) at 1 $\msol$, $({dn\over dm})_{1}= 1.9\,\times 10^{-2}\,\mmvol$, above which the PDMF and the IMF start to differ appreciably ($>10\%$).
This IMF is very similar to the IMF2 derived in Chabrier (2001), which gives a good description of the star counts in the deep field of
the ESO Imaging Survey (EIS) (Groenewegen et al.  2002)
and whose predictions in the BD domain
agree fairly well with present detections of various field surveys (Chabrier 2002).

As demonstrated by the detailed study of Kroupa, Tout \& Gilmore (1993) and Kroupa (1995),
most of the discrepancy between photometric and nearby LFs
for $M_V>12$ results from Malmquist bias and unresolved binary systems in the low-spatial resolution photographic surveys.
Although the Malmquist bias is negligible for the HST,
this latter, however,
misses essentially all companions of multiple systems, because of its angular resolution. GBF97 estimate that the correction arising
from unresolved companions is at most a factor of 2 at $0.1\ \msol$, whereas the difference between
the HST (GBF97) MF and the one derived from $\Phi_{near}$ is more than a factor of 4 in this region
(see e.g. Figure 1 of M\'era et al. 1998 or Figure \ref{fig2_col.ps} below, empty triangles).
Clearly, the binary correction can not account {\it by itself} for the difference.
A major caveat of any photometric LF, however, is that the determination of the distance relies on a photometric determination from a color-magnitude diagram. The former analysis of the HST data (GBF97)
used for the entire sample a color-magnitude transformation characteristic of
stars with solar abundances.
As shown in Figure 2  of Zheng et al. (2001), however,
the vast majority of the stars in the HST sample lie at a Galactic height
$|z| \simgr $ 800 pc above the plane. These stars are expected to
have metal-depleted abundances and fainter magnitudes
for a given V-I color than stars with solar abundance (Chabrier \& Baraffe 2000).
Assuming a solar abundance for the entire HST sample thus results in an overestimation
of the distance and an underestimation of the number density. 
This point was considered recently in the new analysis and sample
of Zheng et al. (2001), yielding a revised $\Phi_{HST}$,
with indeed a larger number of M dwarfs at dim absolute magnitudes. This new sample, however,
does not include the correction due to unresolved binaries and the inferred IMF still differs
significantly from the one derived from the local sample.
We have conducted a detailed analysis of this bias with this new LF.

\section{Binary correction to the local and HST luminosity functions}

\subsection{Analysis of the mass-ratio distribution}

Although the multiplicity rate for {\it stellar} companions of M dwarfs still remains ill-determined,
a reasonable estimate is starting to emerge, with a value $X_\star\approx 30\pm 5$ \% (Marchal et al.  2002).
Mass ratios of binaries have been determined accurately only for  F and G stars
(Duquennoy \& Mayor 1991; DM91). Similar M-dwarf studies are in progress (Delfosse et al. 1999; Marchal et al.  2002), but extended observations
($\sim$ 10 years) are required to get unbiased results.
 The studies conducted by Mazeh et al. (1996),
restricted to short period binaries, give a linear fit of mass-ratio
distribution whose slope is compatible with 0, the uncertainty being
large. The recent determinations by Marchal et al. (2002) point to a mass ratio close to unity for short
period binaries ($P<100$ d), but a distribution compatible with a DM91 or a uniform one for longer periods.

We have conducted Monte Carlo simulations in order to estimate the effect of such unresolved binaries on the local and HST MF.
The mass of the single stars and primaries $m_p$ is drawn randomly according to MF (eq. [\ref{IMFdisk}]).
A fraction $X$ of these stars are then selected with a uniform probability distribution and are attributed a companion. The mass
of the companion is drawn from a mass fraction distribution $P(q)$ ($q$=$m_2/m_1 \le 1$), assuming this distribution does not depend on the mass of the primary.
In order to estimate the dependence of the binary correction upon the parameters, we have conducted calculations
with several binary fractions and mass-ratio distributions, namely
$P(q)$=constant,  $P(q)\propto exp\Bigl(-{(q-\mu)^2\over 2\sigma_q^2}\Bigr)$, $P(q)\propto q$ and $P(q)\propto (1-q)$. These distributions correspond to a uniform mass-ratio distribution, a DM91 distribution for $\mu=0.23$, $\sigma_q=0.42$, and distributions
biased towards equal masses and low mass ratio, respectively. The resulting distribution
$dN=N_{tot}/N_p$, i.e. the total number of stars $N_{tot}=N_p+N_s$ over the number of primaries
increases with decreasing mass approximately as $dN\propto m^{-0.16}$ from $\sim 0.5$ to 0.1$\,\msol$,
with a maximum of $\sim (30\pm 10)\%$ at $m=0.1\,\msol$, for $X=0.5\pm 0.1$.
The shape of the correction is found to depend only weakly on the $P(q)$ distribution.

Notice that these distributions imply that a fraction of the companions are below the H-burning limit ($m<0.072\,\msol$). For the sample studied by DM91, about 60\% of the observed stars have
a companion of mass larger than 0.1 $\msol$
and the DM91 distribution predicts $\sim$10\% of sub-stellar companions. If the same
distribution is applied to a 0.2 $\msol$ M dwarf, then about 50\% of the companions are BDs. This means that the observed
{\it frequency} of {\it stellar} binaries depends on the mass of the primary. In a sample including stellar objects only,
the present calculations predict an {\it observable} (stellar) fraction of companions $\sim$ 60\% among M-dwarf
primaries, the remaining $\sim$ 40\% fraction being BD companions. For a $X=$50\% binary frequency (see below), this
implies $\sim 30\%$ of M-dwarf M-dwarf systems, and $\sim 20\%$ of systems composed
of an M dwarf with a BD companion. This is in good agreement with the presently observed M dwarf binary fraction in the solar neighborhood
(Delfosse et al. 1999; Marchal et al. 2002), and with the present estimates of BD companions of M dwarfs at large orbital
separations (Gizis et al. 2001).
The correction to the LF and to the MF
is examined below.

\subsection{Effect of binary correction to the luminosity function and mass function}

We first consider the effect of unresolved binaries on the MF derived from the nearby LF $\Phi_{near}$.
For that, we have merged the identified companions in the Dahn et al. (1986) and Henry \& McCarthy
(1990) samples into unresolved systems\footnote{For the Henry \& McCarthy sample, we have also merged the
binaries GL 15 A and B into one system to get the complete system LF from their Fig. 10b}.
This yields the nearby {\it system} LF, from which we have calculated the {\it system} MF, following the
same procedure as in \S2.
Figure \ref{fig2_col.ps} displays this system MF, as well as the recent HST MF (Table 4 of Zheng et al.  2001). 
The two MFs are compatible at the $< 1\sigma$ level.
For comparison, the figure displays also the HST MF obtained from a color-magnitude distance determination
assuming that all the objects have a solar abundance, as done in GBF97 (Zheng et al.  2001, Figure 4 with their CMR(1)).
This latter is much more difficult to reconcile with the local system MF below $m\la 0.25\,\msol$, as mentioned earlier.
For further purposes, it is interesting to parametrize this {\it system} MF, as done in equation [\ref{IMFdisk}]
for the single objects, as:

\begin{eqnarray}
\xi(\log \,m)_{sys}={dn \over d\log \,m}=0.086\times \exp\{-{(\log \, m\,\,-\,\,\log \, 0.22)^2\over 2\times (0.57)^2}\}\,\,\,\,{\rm pc}^{-3}\,(\log \,\msol)^{-1}
\label{IMFsys}
\end{eqnarray}

\noindent with the same normalization as MF (eq. [\ref{IMFdisk}]) at 1 $\msol$, where the binary correction is negligible.
It is displayed by the solid line in Figure \ref{fig2_col.ps}.

In order to verify this correction due to unresolved binaries on the local LF $\Phi_{near}$, we have applied
the same type of Monte Carlo simulations as described above. However, in the present case, we have explored the
possibility that the primary and the secondary are
drawn randomly from the {\it same } single object MF (eq. [\ref{IMFdisk}]).
The {\it system} LF is then calculated by attributing a
magnitude $M_{sys} = -2.5\,\log[10^{-0.4\,M(m_1)}+10^{-0.4\,M(m_2)}]$ to the unresolved binary and the system MF is derived
with the same MMRs as in \S2. The resulting system MF is displayed in
Figure \ref{fig3_col.ps} for a binary fraction $X=50\%$ (solid line) and
$X=30\%$ (dash line). As seen in the figure, the agreement with the observed
local system MF is excellent, and the system MF for $X=50\%$ agrees
surprisingly well with the parametrized form (eq. [\ref{IMFsys}]).

The quantification of the effect of unresolved binaries on the HST MF is more
complicated, for in that case the Galactic scale height variation must be taken into account.
We use the same Monte-Carlo calculations, with the disk density profile $\rho(R,z)$ determined by Zheng et al. (2001, eq. [4]).
We then proceed exactly as for the local LF, with
a simulated stellar population, including $X\%$ binary systems,  drawn
randomly from this spatial distribution, with masses given by equation [\ref{IMFdisk}]. We use the same $\mv$-$(V-I)$ relation and color cut 1.53$<V-I<$4.63 as Zheng et al. (2001). We then
reconstruct the HST observed LF obtained with the $1/V_{\rm max}$ method ($\Phi=\Sigma (N/V_{max})$), assuming all binary systems unresolved. The reconstructed MF from this system LF
is compared to the one derived by Zheng et al. (2001, squares) on Figure (\ref{fig3_col.ps}) for
$X=50\%$ (dash-dot line).
The HST data have been multiplied by a factor $7.1/8.1$ to bring the HST normalization at 0.6 $\msol$
into consistency with the one inferred from equation [\ref{IMFdisk}] (see Zheng et al. 2001, \S3.3).
The simulated HST MF including the effect of unresolved companions is consistent at the $<2\sigma$ level with the observed one, the remaining discrepancy arising most likely from the MMR metallicity-dependent correction used in the HST sample analysis or from incompleteness of this sample at dim magnitudes.
Surprisingly, the main difference between the reconstructed
HST system MF and the local system MF  (eq.[\ref{IMFsys}])
occurs for the larger masses ($m\ga 0.4\,\msol$). The reason is the
Malmquist bias in the $1/V_{max}$ method used in the present simulations and in GBF97
due to the saturation threshold of the HST camera, $I_{min}=18.75$, which excludes a non negligible fraction of the simulated stars. This bias, however, is corrected in the
maximum likelihood analysis done by Zheng et al. (2001).
The simulations for the {\it volume}-limited
local sample are not affected by this bias and yield agreement between
the simulated and system MF (\ref{IMFsys}) over the entire considered
mass range.

\section{Conclusion}
\label{Conclusion}

In this Letter, we have derived the single and systemic MFs for the Galactic disk in the M-dwarf regime,
from both the V and K-band local LFs.
Both determinations are well reproduced by a lognormal form,
normalized at 1 $\msol$ on the value
derived by Scalo (1986).
We have shown 
that the disk stellar MF determined from either the nearby parallax LF or the HST photometric LF are consistent, and that the previous source of disagreement was due to {\it two} cumulative effects, namely (i) incorrect color-magnitude determined distances in the original LF derived by GBF97, due to the fact that a large fraction of the HST M-dwarf sample belongs to a metal-depleted
population high above the Galactic plane, a point corrected in the recent analysis of Zheng et al. (2001) and (ii) unresolved binaries in the HST sample\footnote{Multiple systems besides binaries will bring further correction.
This latter, however, is likely to be small. Indeed, out of the known 39 M dwarfs within 5 pc,
8 belong to binaries, but only 2 to a triple system (Henry \& McCarthy 1990).}.
We have shown that the HST MF is very similar to
the local {\it system} MF. This latter is consistent with a fraction $X\sim $50\% of binaries, with masses for the primary and
the companions determined by the same underlying aforementioned single MF. This
yields roughly equal fractions of M-dwarf M-dwarf and M-dwarf BD systems,
in agreement with present observations. 

These results yield a reinterpretation of the so-called "brown-dwarf desert".
This latter, expressing the deficit of small-separation BD companions to solar-type stars,
as compared with stellar or planetary companions, has sometimes been interpreted as an IMF of substellar companions to solar-type stars significantly different from the one determined for the field.
The present calculations, however, show that this "desert" should be reinterpreted as a lack of high mass-ratio ($q\la 0.1/1$)
systems, and does not preclude a substantial fraction of BDs as companions
of M dwarfs or other BDs. Moreover, BD companions of stars, i.e. systems with large mass-ratio, may form preferentially at large separations, requiring long time basis for detection, as suggested by the recent analysis of Marchal et al. (2002).
The present calculations and the ones developed in Chabrier (2001, 2002) suggest that stars,
BDs and
companions originate from the same universal IMF (eq. [\ref{IMFdisk}]).

\acknowledgments
The author is endebted to the referee, Andy Gould, for helping improving the original manuscript.

\vfill\eject

\eject

\begin{figure}
\centerline{\it Figure Legends}

\caption[]{Disk IMF derived from the local V-band LF (circles and solid line)
and K-band LF (squares and dash-line).
The solid line displays the lognormal form 
(eq.[\ref{IMFdisk}]).
}
\label{fig1_col.ps}
\end{figure}

\begin{figure}
\caption[]{Disk MF derived from the system K-band LF (squares and solid line) and the
HST corrected MF (solid triangles and dash-line) from Zheng et al. (2001). 
 The Zheng et al. (2001) MF has been
multiplied by a factor $7.1/8.1$ to bring the HST normalization at 0.6 $\msol$
consistent with the one inferred from eq.[\ref{IMFdisk}].
 The solid line and upper dot-dash-line illustrate
the lognormal form given by eq. [\ref{IMFsys}] and eq. [\ref{IMFdisk}], respectively. 
The HST MF obtained if all objects are assumed to have a solar metallicity (see Zheng et al. 2001)
is illustrated by the empty triangles.
}
\label{fig2_col.ps}
\end{figure}

\begin{figure}
\caption[]{Effect of unresolved binaries on the local and HST MFs.
Circles : nearby {\it system} MF;
squares: HST MF corrected for metallicity gradient, as in Figure 2.
Solid and dashed lines: reconstructed local {\it system} MF, for
50\% (solid curve) and 30\% (dash curve) of unresolved binaries, respectively;
dot-dash line: reconstructed HST {\it system} MF for 50\% of unresolved binaries.
Upper dotted line:
single object IMF (eq.[\ref{IMFdisk}]); lower dotted line: system IMF (eq.[\ref{IMFsys}]).}
\label{fig3_col.ps}
\end{figure}

\vfill\eject

\begin{figure}
\begin{center}
\epsfxsize=180mm
\epsfysize=180mm
\epsfbox{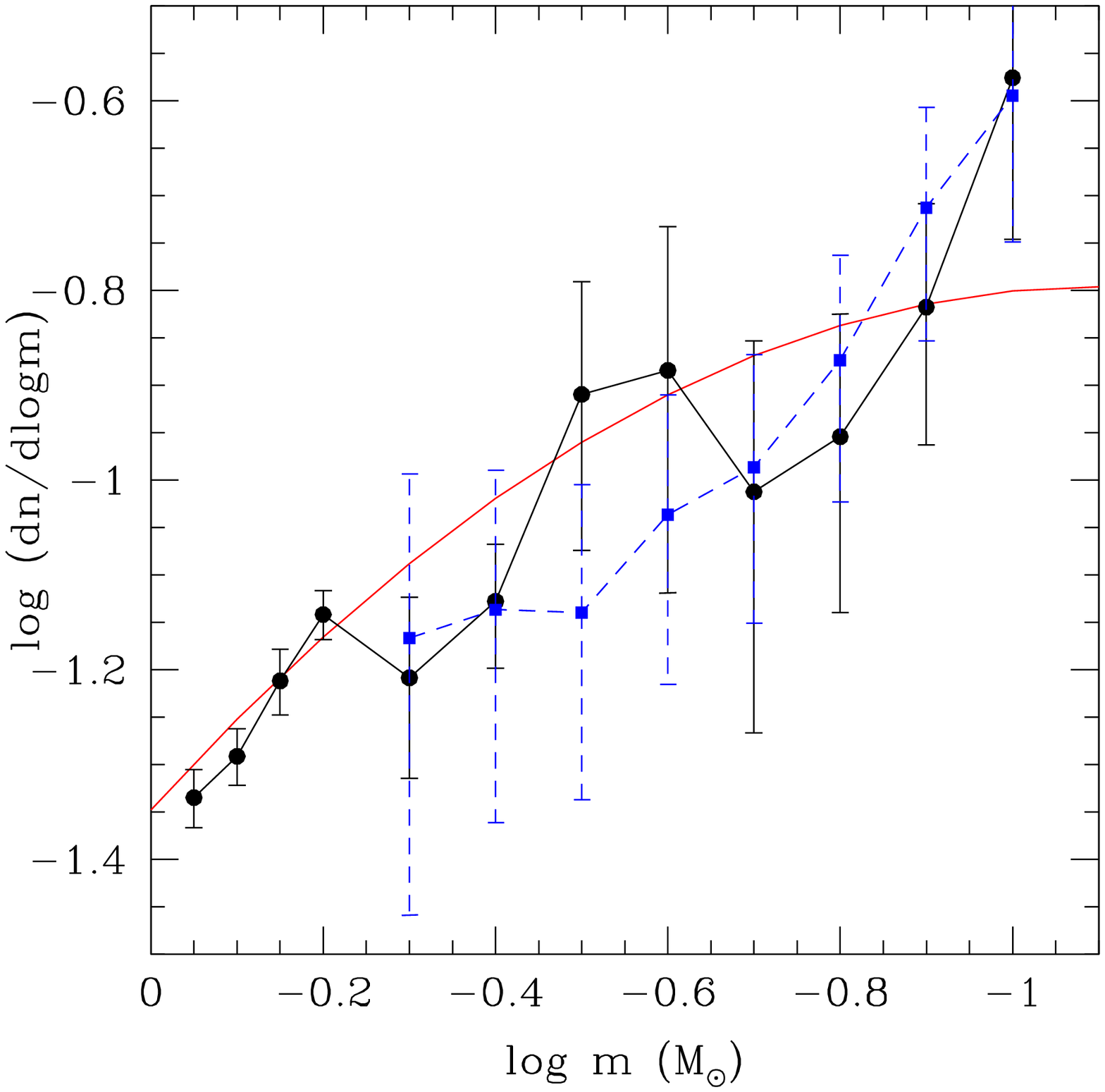}
\end{center}
\end{figure}

\vfill\eject

\begin{figure}
\begin{center}
\epsfxsize=180mm
\epsfysize=180mm
\epsfbox{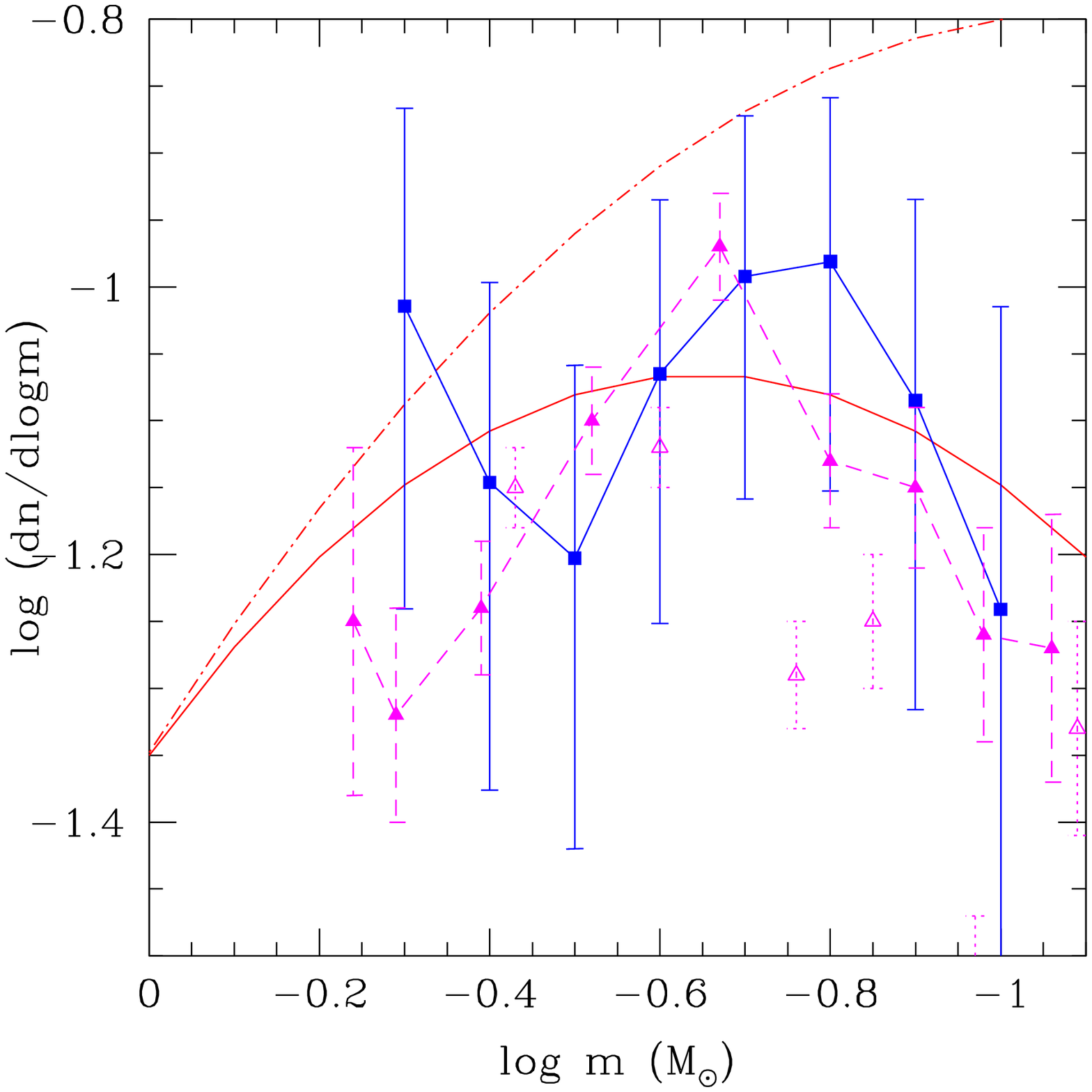}
\end{center}
\end{figure}

\vfill\eject

\begin{figure}
\begin{center}
\epsfxsize=180mm
\epsfysize=200mm
\epsfbox{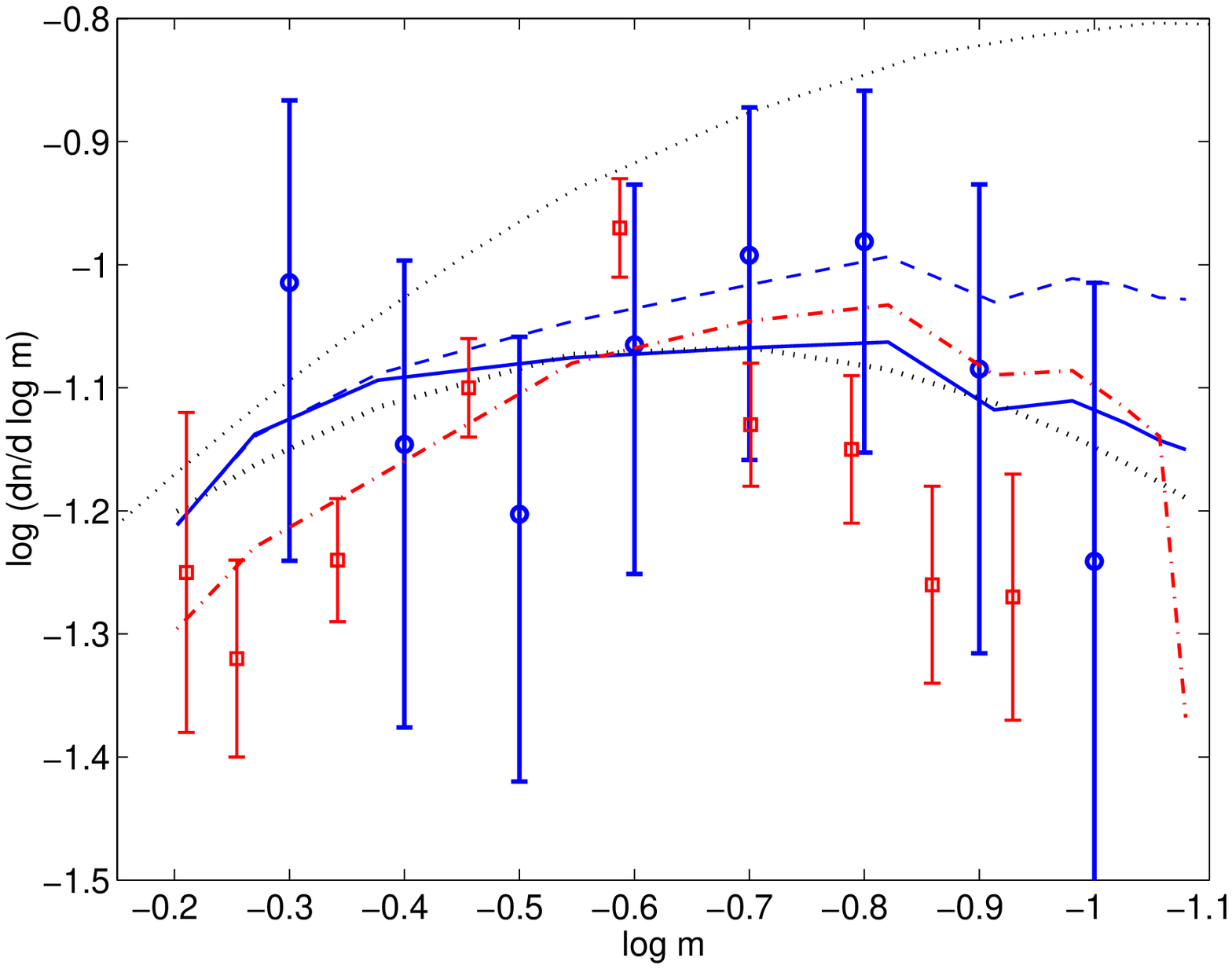}
\end{center}
\end{figure}

\end{document}